\documentclass[12pt,a4,onecolumn,secnumarabic,amssymb, amsmath, nofootinbib,tightenlines,
nobibnotes, aps, prl,epsfig]{revtex4}
\usepackage{graphicx}
\usepackage{dcolumn}
\usepackage{bm}
\begin{document}
\preprint{APS/123-QED}
\title{The approximation method for calculation of the exponent of the gluon distribution-$\lambda_{g}$
 and the structure function-$\lambda_{S}$ at low $x$}

\author{G.R.Boroun }
\altaffiliation{boroun@razi.ac.ir}
\author{B.Rezaie}%
\affiliation{ Physics Department, Razi University, Kermanshah
67149, Iran}
\date{\today}

\begin{abstract}
We present a set of formulae using the solution of the QCD
Dokshitzer-Gribov-Lipatov-Altarelli-parisi (DGLAP) evolution
equation to the extract of the exponent $\lambda_g$ gluon
distribution and $\lambda_S$ structure function from the Regge-
like behavior at low $x$. The exponents are found to be
 independent of $x$ and to increase linearly with ln$Q^{2}$ and compared with the most data
  from H1 Collaboration. We also calculated  the structure function
  $F_{2}(x,Q^{2})$ and the gluon distribution $G(x,Q^{2})$ at low
  $x$ assuming the Regge- like behavior of the gluon distribution
  function at this limit and compared with NLO QCD fit to the H$1$
  data, two Pomeron fit, multipole Pomeron exchange fit and MRST
  (A.D.Martin, R.G.Roberts, W.J.Stirling and R.S.Thorne), DL(A.Donnachie and
  P.V.Landshoff), NLO-GRV(M.Gluk, E.Reya and A.Vogt) fit results,
  respectively.

\end{abstract}

\maketitle
\subsection{1 Introduction}
The Knowledge of the deep inelastic scattering (DIS) structure
functions at small values of the Bjorken scaling variable $x$ is
interesting for understanding the inner structure of hadrons. Of
great relevance is the determination of the gluon density at low-
$x$, where gluons are expected to be dominant, because it could be
a test of perturbative quantum chromodynamic (PQCD) or a probe of
new effects, and also because it is the basic ingredient in many
other calculations of different high energy hadronic processes.\\

The behavior of the proton structure function $F_{2}(x,Q^{2})$ at
small $x$ reflects the behavior of the gluon distribution, since
the gluon is by far the dominant parton in this regime. At small
$x$, only the structure function $F_{2}$ is measured. On the other
hand, the gluon distribution cannot be measured directly from
experiments. It is, therefore, important to measure the gluon
distribution $G(x,Q^{2})$ indirectly from the proton structure
function $F_{2}(x,Q^{2})$ through the transition
$g{\rightarrow}\hspace{0.1cm}q\overline{q}$. Here the
representation for the gluon distribution $G(x)=xg(x)$ is used,
where $g(x)$ is the gluon
density.\\

In PQCD, the high- $Q^{2}$ behavior of DIS is given by the
Dokshitzer- Gribov- Lipatov- Altarelli- Parisi (DGLAP) evolution
equations [1]. In the double asymptotic limit (large energies,
i.e. small- $x$ and large photon virtualities $Q^{2}$), the DGLAP
evolution equations can be solved [2] and structure function is
expected to rise approximately  like a power of $x$ towards small-
$x$. This steep rise of $F_{2}(x,Q^{2})$ towards low- $x$ observed
at HERA [3], also indicates in PQCD a similar rise of the gluon
towards low- $x$. This similar behavior predicts a steep power law
behavior for gluon distribution. Accordingly the approximate
solutions of DGLAP evolution equations are reported in recent
years [4,5] with
considerable phenomenological success.\\

The small- $x$ region of DIS offerees a unique possibility to
explore the Regge limit of PQCD [6,7]. This theory is successfully
described by the exchange of a particle with appropriate quantum
numbers and the exchange particle is called a Regge pole.
Phenomenologically, the Regge pole approach to deep inelastic
scattering implies that the structure functions are sums of powers
in $x$, modulus logarithmic terms, each with a $Q^{2}$- dependent
residue factor. This model gives the following parametrization of
the deep inelastic scattering structure function $F_{2}(x,Q^{2})$
at small $x$,
$F_{2}(x,Q^{2})=\sum_{i}A_{i}(Q^{2})x^{-\lambda_{i}}$, that the
singlet part of the structure function is controlled by pomeron
exchange at small $x$. The rapid rise in $Q^{2}$ of the structure
functions was considered as a sign of departure from the standard
Regge behavior. The reason was that the HERA data, when fitted by
a single "Regge- pomeron" term ${\sim}\hspace{0.1cm}
x^{-\lambda_{S}}$, where $\lambda_{S}$ is the pomeron intercept
minus one, show that
$\lambda_{S}=\frac{d{\ln}F_{2}(x,Q^{2})}{d{\ln}\frac{1}{x}}$
definitely
rises with $Q^{2}$.\\

The Regge behavior of the structure function in the large- $Q^{2}$
region reflects itself in the small- $x$ behavior of the quark and
the antiquark distributions. Thus the Regge behavior of the sea
quark and antiquark  distribution for small- $x$ is given by
$q_{sea}(x){\sim}\hspace{0.1cm}x^{-\alpha_{p}}$ corresponding to a
pomeron exchange with an intercept of $\alpha_{p}=1$. But the
valance quark distribution for small- $x$ given by
$q_{val}(x){\sim}\hspace{0.1cm}x^{-\alpha_{R}}$ corresponds to a
Reggeon exchange with an intercept of $ \alpha_{R}=1/2$. The small
$x$ behavior of the structure functions is driven by gluon through
the $g{\rightarrow}q\overline{q}$ transition and the increase of
gluon distributions with decreasing $x$ implies a similar increase
of the deep inelastic lepton- proton scattering structure function
$F_{2}$ as the Bjorken parameter $x$ decreases, so we expect
$g(x){\sim}1/x$. The $x$- dependence of the parton
densities given above is often assumed at moderate- $Q^{2}$.\\

In principle, the HERA data should determine the small $x$
behavior of gluon and sea quark distribution. Roughly speaking the
data on the singlet part of the structure function $F_{2}$
constrain the sea quarks and the data on the slope
$dF_{2}$/dln$Q^{2}$ determine the gluon density. For example if we
take:

\begin{equation}
F_{2}{\sim}\hspace{0.1cm}xS \hspace{0.1cm}{\sim}\hspace{0.1cm}
A_{S}x^{-\lambda_{S}},
\end{equation}
and
\begin{equation}
\frac{dF_{2}}{d{\ln}Q^{2}}\hspace{0.1cm}{\sim}\hspace{0.1cm}xg
{\sim}\hspace{0.1cm} A_{g}x^{-\lambda_{g}},
\end{equation}
then we might expect to determine $\lambda_{S}$ and $\lambda_{g}$.
In the DGLAP formalism the gluon splitting functions are singular
as $x{\rightarrow}\hspace{0.1cm}0$. Thus the gluon distribution
will become large as $x{\rightarrow}\hspace{0.1cm}0$, and its
contribution to the evolution of the parton distribution becomes
dominant. In particular the gluon will drive the quark singlet
distribution, and hence the structure function $F_{2}$ becomes
large as well, the rise
increasing in steepness as $Q^{2}$ increases.\\

The rapid rise in $Q^{2}$ of the gluon distribution at small $x$,
observed at HERA, shows by[8]:
\begin{equation}
xg(x,Q^{2})=\hspace{0.1cm} A_{g}x^{-\lambda_{g}},
\end{equation}
where $\lambda_{g}$ is the pomeron intercept minus one. In the
double asymptotic limit (large energies. i.e. small $x$, and large
photon virtualities $Q^{2}$) the DGALP evolution equations [1] can
be solved [2] and $F_{2}$ is expected to rise approximately like a
power of $x$ towards low $x$. In leading order the DGLAP gluon
distribution solution gives [2]:
\begin{equation}
\lambda_{g}=[\frac{12}{\beta_{0}}\frac{\ln(\frac{t}{t_{0}})}{\ln(\frac{1}{x})}]^{1/2},
\end{equation}
where t=ln$(\frac{Q^{2}}{\Lambda^{2}})$,
$t_{0}$=ln$(\frac{Q_{0}^{2}}{\Lambda^{2}})$ (that $\Lambda$ is the
QCD cut- off parameter) and $\beta_{0}=\frac{33-2N_{f}}{3}$,
$N_{f}$ being the number of flavours. This steep behavior of the
gluon generates a similar steep behavior of $F_{2}$ at small $x$,
Eq.(1), where $\lambda_{S}=\lambda_{g}-\epsilon$ (i.e.
$\lambda_{S}{\neq}\lambda_{g}$ in next- to- leading order
analysis). At small $x$ terms in ln$(1/x)$ are becoming large and
the conventional leading ln$Q^{2}$ summation of the DGLAP
equations does not account for this. It may  also be necessary to
sum leading ln$(1/x)$ terms. Such a summation is performed by the
BFKL [9] equation. To leading order in ln$(1/x)$ with fixed
$\alpha_{s}$, this predicts a steep power law behavior
$xg(x,Q^{2}){\sim}x^{-\lambda_{g}}$ where
$\lambda_{g}=\frac{3\alpha_{s}}{\pi}4\ln2{\simeq}0.5$ (for
$\alpha_{s}{\simeq}0.2$, as appropriate for
$Q^{2}{\sim}\hspace{0.1cm}4 $GeV$^{2}$
).\\

For $Q^{2}{\leq}1\hspace{0.1cm}$GeV$^{2}$, the simplest Regge
phenomenology predicts that the value of
$\lambda_{S}=\alpha_{\mathbb{P}}(0)-1{\simeq}\hspace{0.1cm}0.08$
is consistent with that of hadronic Regge theory [3,10] where
$\alpha_{\mathbb{P}}(0)$ is described soft pomeron dominant with
its intercept slightly above unity ($\sim$1.08), whereas for
$Q^{2}{\geq}1\hspace{0.1cm}$GeV$^{2}$ the slope rises steadily to
reach a value greater than $0.3$ by
$Q^{2}{\approx}100\hspace{0.1cm}$GeV$^{2}$ where hard pomeron is
dominant. This larger value of $\lambda_{S}$ is not so far from
that expected using BFKL [9] ideas. Indeed, the BFKL equation can
be viewed as a method for the calculation of the hard or
perturbative Pomeron trajectory (in contrast to the soft or non-
perturbative Pomeron of hadronic physics with intercept around
$1.08$ [11]). There is some  other authors [10] which extended
their Regge model adding a hard Pomeron with intercept $1.44$,
which allows them to describe the low $x$ HERA data up to $Q^{2}$
values of a few hundred
GeV$^{2}$.\\

Our goal in this work is to present an approximate analytical
solution for the singlet structure function and the gluon
distribution as we should be able to calculate $\lambda_{S}$ and
$\lambda_{g}$ in the next- to- leading order DGLAP equation, valid
to be at low- $x$. In order to do this, the DGLAP evolution
equations are calculated  neglecting the quark distribution. The
approach, using the Regge  and the Regge- like behavior for
singlet and gluon distribution respectively, has been applied in
this paper. We test its validity comparing it with that of $H1$
Collaboration  and attempt to see how the predictions for singlet
exponent are compared with the  experimental data [3].\\

The formulation of the problem in NLO-DGLAP evolution equations
for calculation $\lambda_{g}$ and $\lambda_{S}$  are given in
Sections.2 and 3. Finally in Secion.4, the numerical results are
given,
leading to discussion and conclusions.\\

\subsection{2. Calculation of $\lambda_{g}$ based on  Regge- like behavior gluon distribution function  }

Neglecting the quark singlet part, the DGLAP equation for the
gluon evolution in the NLO can be written as [12]:
\begin{equation}
Q^{2}\frac{{\partial}G}{{\partial}Q^{2}}=\frac{\alpha_{s}}{2\pi}{\int_{x}^{1}}
[P^{1}_{gg}(z)+\frac{\alpha_{s}}{2\pi}P^{2}_{gg}(z)]
G(\frac{x}{z},Q^{2})dz,
\end{equation}
where $P^{1}_{gg}(z)$ and $P^{2}_{gg}(z)$ are the LO and NLO
Altarelli- Parisi splitting kernels [1,12]. The running coupling
constant $\alpha_{s}(Q^{2})$ has the approximate analytical form
in NLO:
\begin{equation}
\frac{\alpha_{s}(Q^{2})}{2\pi}=\frac{2}{\beta_{0}\ln(\frac{Q^{2}}{\Lambda^{2}})}
[1-\frac{\beta_{1}\ln\ln(\frac{Q^{2}}{\Lambda^{2}})}{\beta_{0}^{2}\ln(\frac{Q^{2}}{\Lambda^{2}})}],
\end{equation}
where  $\beta_{0}=\frac{1}{3}(33-2N_{f})$ and
$\beta_{1}=102-\frac{38}{3}N_{f}$ are the one- loop (LO) and the
two- loop (NLO) correction to the QCD $\beta$- function, $N_{f}$
being the number of active quark flavours ($N_{f}=4$). To find an
analytic solution, we note that the splitting kernels
$P^{1}_{gg}(x)$ and $P^{2}_{gg}(x)$ at the small $x$ limit are
[13,14]:
\begin{eqnarray}
P^{1}_{gg}(x)=2C_{A}[\frac{x}{(1-x)_{+}}+\frac{1-x}{x}+x(1-x)]\\\nonumber
+{\delta}(1-x)\frac{(11C_{A}-4N_{f}T_{R})}{6},
\end{eqnarray}
\begin{eqnarray}
P^{2}_{gg}(x)=\frac{(12C_{F}N_{f}T_{R}-46C_{A}N_{f}T_{R})}{9x}+N_{f}T_{R}(\frac{-61}{9}C_{F}\nonumber\\
+\frac{172}{72}C_{A})+C^{2}_{A}(\frac{1643}{54}-\frac{22}{3}\zeta(2)-8\zeta(3)).\nonumber\\
\end{eqnarray}
Where the casimir operators of colour SU(3) are defined as:
\begin{eqnarray}
C_{A}=3, \hspace{0.5cm}C_{F}=\frac{4}{3},
\hspace{0.5cm}T_{R}=\frac{1}{2},
\end{eqnarray}
and $[f(x)]_{+}{\equiv}f(x)-\delta(1-x)\int_{0}^{1}f(y)dy$. Using
Eqs.(7-9) in Eq.(5) and carrying out the integration we get:
\begin{equation}
\frac{dG(x,t)}{dt}=\frac{3\alpha_{s}}{\pi}I_{\lambda_{g}}
G(x,t)+(\frac{\alpha_{s}}{2\pi})^{2}T_{\lambda_{g}} G(x,t),
\end{equation}
where
\begin{eqnarray}
T_{\lambda_{g}}&=&\frac{(12C_{F}N_{f}T_{R}-46C_{A}N_{f}T_{R})}{9\lambda_{g}}(1-x^{\lambda_{g}})\nonumber\\
&&+[N_{f}T_{R}(\frac{-61}{9}C_{F}+\frac{172}{72}C_{A})+C^{2}_{A}(\frac{1643}{54}-\frac{22}{3}\zeta(2)\nonumber\\
&&-8\zeta)3))]\frac{1-x^{1+\lambda_{g}}}{1+\lambda_{g}},
\end{eqnarray}
\begin{eqnarray}
I_{\lambda_{g}}&=&(\frac{11}{12}-\frac{N_{f}}{18})+ln(1-x)+{\int_{x}^{1}}dz[\frac{z^{1+\lambda_{g}}-1}{1-z}\nonumber\\
&&+(1-z)(z^{1+\lambda_{g}}+z^{\lambda_{g}-1})].
\end{eqnarray}
and
\begin{eqnarray}
{\int_{x}^{1}}dz[\frac{z^{1+\lambda_{g}}-1}{1-z}+(1-z)(z^{1+\lambda_{g}}+z^{\lambda_{g}-1})]\nonumber\\
=\frac{2}{2+\lambda_{g}}(1-x^{2+\lambda_{g}})-(1-x)-\frac{1}{2}(1-x^{2})\nonumber\\
+\frac{1}{\lambda_{g}}(1-x^{\lambda_{g}})-\frac{1}{1+\lambda_{g}}(1-x^{1+\lambda_{g}})\nonumber\\
+\sum_{N=4}^{\infty}[\frac{1}{N+\lambda_{g}}(1-x^{N+\lambda_{g}})-\frac{1}{N-1}(1-x^{N-1})].
\end{eqnarray}
Eq.(10) can be rearranged as:
\begin{equation}
\frac{dG(x,t)}{dt}=(\frac{3\alpha_{s}}{\pi}I_{\lambda_{g}}
+\frac{\alpha_{s}^{2}}{4\pi^{2}}T_{\lambda_{g}}) G(x,t).
\end{equation}\\
We note that exponent $\lambda_{g}$ is given as the derivative
\begin{equation}
\lambda_{g}=\frac{{\partial}{\ln}G(x,t)}{{\partial}{\ln}\frac{1}{x}}|_{t=constant}.
\end{equation}\\
To obtain an expression for $\lambda_{g}$ we first differentiate
Eq.(14) with respect to ln$(1/x)$ and then integrate from $t_{0}$
to ${t}$. Finally, as $x{\rightarrow}\hspace{0.1cm}0$, we
retaining only its leading terms, i.e., the approximate analytical
solution is given as follows:
\begin{equation}
\lambda_{g}G(x,t)-\lambda_{g_{0}}G(x,t_{0})={\int_{t_{0}}^{t}}G(x,t)(\frac{3\alpha}{\pi}-\frac{61\alpha^{2}}{9\pi^{2}})dt.
\end{equation}
where
$\lambda_{g_{0}}$(=$\frac{{\partial}{\ln}G(x,t_{0})}{{\partial}{\ln}\frac{1}{x}}$)
is the exponent at the starting scale $t_{0}$  while $G(x,t_{0})$
is the input gluon distribution. On the other hand, Eq.(14) in
low- $x$ has the explicit form:
\begin{equation}
\ln\frac{G(x,t)}{G(x,t_{0})}={\int_{t_{0}}^{t}}(\frac{3\alpha}{\pi}-\frac{61\alpha^{2}}{9\pi^{2}})\frac{1-x^{\lambda_{g}}}{\lambda_{g}}dt.
\end{equation}
Hence, we obtain an approximation expression for $\lambda_{g}$ as
\begin{eqnarray}
\ln\frac{\lambda_{g_{0}}}{\lambda_{g}-x^{\lambda_{g}}{\int_{t_{0}}^{t}}x^{-\lambda_{g}}(\frac{3\alpha}{\pi}-\frac{61\alpha^{2}}{9\pi^{2}})dt}\nonumber\\
={\int_{t_{0}}^{t}}(\frac{3\alpha}{\pi}-\frac{61\alpha^{2}}{9\pi^{2}})\frac{1-x^{\lambda_{g}}}{\lambda_{g}}dt.
\end{eqnarray}\\

The low $x$ behavior of $F_{2}$ at fixed $Q^{2}$ is studied
locally by the measurement of the derivative
$\lambda_{S}{\equiv}-({\partial}{\ln}F_{2}/{\partial}{\ln}x)_{Q^{2}}$
as function of $x$ and $Q^{2}$ [3]. The new shifted vertex and the
published data agree well in the overlap region. The derivative
$\lambda(x,Q^{2})$ is independent of $x$ for $x{<}0.01$ to within
the experimental accuracy. This implies that $x$ dependence of
$F_{2}$ at low $x$ is consistent with a power law
$F_{2}{\sim}x^{-\lambda_{S}(Q^{2})}$, for fixed $Q^{2}$. As,
$\lambda_{S}(Q^{2})$ rises approximately linearly with $lnQ^{2}$.
The rise of the proton structure function towards small $x$ has
been discussed since the gluon distribution is dominated within
the proton at low $x$. So, the function $\lambda_{g}$ also rises
approximately linearly with ln$Q^{2}$. This dependence  can be
represented as $\lambda_{g}(Q^{2})=b_{g}\ln[Q^{2}/\Lambda^{2}]$
where $b_{g}$ is a constant. Hence, after the determination of
$\lambda_{g}$, we can calculate gluon distribution function from
Eq.(17) and compare our results with those of other authors. We
take all our inputs from NLO- GRV[15] and $H1$ collaboration [16] and MRST2001[17].\\

\subsection{3. Calculation of $\lambda_{S}$ based on Regge-like behavior of the structure function }

The main task is the determination of singlet exponent  structure
function from the NLO- DGLAP evolution equations. As it can be
seen, for small- $x$, the gluon term, dominates over the scaling
violation of $F_{2}$. Neglecting the quark, the DGLAP evolution
equation for the singlet structure function has the form:
\begin{equation}
\frac{dF^{S}_{2}}{dt}=\frac{\alpha_{s}}{2\pi}{\int_{x}^{1}}dz
(2N_{f}P_{qg}^{1}(z)+\frac{\alpha_{s}}{2\pi}P_{qg}^{2}(z))G(\frac{x}{z},Q^{2}).
\end{equation}
where $P^{1}_{qg}(z)$ and $P^{2}_{qg}(z)$ are the LO and NLO
Altarelli- Parisi splitting kernels [1,12]. The splitting kernels
$P^{1}_{qg}(x)$ and $P^{2}_{qg}(x)$ can be written as [12,13]:
\begin{equation}
P^{1}_{qg}(x)=\frac{1}{2}[x^{2}+(1-x)^{2}],
\end{equation}
and
\begin{eqnarray}
P^{2}_{qg}(x)&=&C_{F}N_{f}T_{R}{\{}4-9x-(1-4x)\ln{x}-(1-2x)\ln^{2}x+4\ln(1-x)+[2\ln^{2}(\frac{1-x}{x})\nonumber\\
&&-4\ln(\frac{1-x}{x})-\frac{2}{3}\pi^{2}+10]P_{qg}(x){\}}+C_{A}N_{f}T_{R}{\{}\frac{182}{9}+\frac{14}{9}x\nonumber\\
&&+\frac{40}{9x}+(\frac{136}{3}x-\frac{38}{3}){\ln}x-4\ln(1-x)-(2+8x)\ln\ln^{2}x+2P_{qg}(-x)S_{2}(x)\nonumber\\
&&+[-\ln^{2}x+\frac{44}{3}{\ln}x-2\ln^{2}(1-x)+4\ln(1-x)+\frac{\pi^{2}}{3}-\frac{218}{9}]P_{qg}(x){\}}
\end{eqnarray}
where $P_{qg}(x)=x^{2}+(1-x)^{2}$ and
$S_{2}(x)={\int_{\frac{x}{1+x}}^{\frac{1}{1+x}}\frac{dz}{z}\ln(\frac{1-z}{z})}$.
 The small- $x$ limit of the next- to-
leading order splitting function for the evolution of the singlet
quark is then [14]:
\begin{equation}
P^{2}_{qg}(x){\longrightarrow}\frac{\alpha_{s}}{2\pi}\frac{40C_{A}N_{f}T_{R}}{9x}.\hspace{3.1cm}
\end{equation}\\

Based on  the Regge- like behavior of the gluon distribution, let
us putting Eq.(3) in Eq.(19). Thus, Eq.(19) is reduced to:
\begin{eqnarray}
\frac{dF^{S}_{2}}{dt}&=&\frac{2\alpha_{s}}{\pi}G(x,t)[\frac{2}{3+\lambda_{g}}(1-x^{{3+\lambda_{g}}})+\frac{1}{1+\lambda_{g}}(1-x^{{1+\lambda_{g}}})-\frac{2}{2+\lambda_{g}}(1-x^{{2+\lambda_{g}}})]\nonumber\\
&&+\frac{120}{18}(\frac{\alpha_{s}}{\pi})^{2}G(x,t)\frac{1-x^{\lambda_{g}}}{\lambda_{g}}.
\end{eqnarray}
or
\begin{eqnarray}
\frac{dF_{2}}{dt}&=&\frac{5\alpha_{s}}{9\pi}G(x,t)[\frac{2}{3+\lambda_{g}}(1-x^{{3+\lambda_{g}}})+\frac{1}{1+\lambda_{g}}(1-x^{{1+\lambda_{g}}})-\frac{2}{2+\lambda_{g}}(1-x^{{2+\lambda_{g}}})]\nonumber\\
&&+\frac{50}{27}(\frac{\alpha_{s}}{\pi})^{2}G(x,t)\frac{1-x^{\lambda_{g}}}{\lambda_{g}}.
\end{eqnarray}

 Therefore, the proton structure function in the low- $x$ region
should be determined from Eq.(24). To continue, we want to
calculate the exponent $\lambda_{S}$ as the derivative
\begin{equation}
\lambda_{S}=\frac{{\partial}{\ln}F_{2}(x,t)}{{\partial}{\ln}\frac{1}{x}}|_{t=cte},
\end{equation}
and compare the prediction with the $H1$ data [3] where the
measurement of the exponent in a large kinematical domain at low
$x$,
$3.10^{-5}\hspace{0.1cm}{\leq}\hspace{0.1cm}x\hspace{0.1cm}{\leq}\hspace{0.1cm}0.2$
and
$1.5\hspace{0.1cm}{\leq}\hspace{0.1cm}Q^{2}\hspace{0.1cm}{\leq}\hspace{0.1cm}150\hspace{0.1cm}
$GeV$^{2}$ has been reported. The exponent $\lambda_{S}$ being
directly measurable from the structure function data can give us
helpful insight into the behavior of the structure function at
low-$x$. In order to find solution for $\lambda_{S}$ we first
differentiate Eq.(24) with respect to ln$1/x$ and then integrate
from $t_{0}$ to $t$. Finally we obtain:
\begin{eqnarray}
\lambda_{S}F_{2}(x,t)-\lambda_{S_{0}}F_{2}(x,t_{0})&=&\frac{0.555}{\pi}{\int_{t_{0}}^{t}}\alpha_{s}G(x,t)[(\frac{2\lambda_{g}}{3+\lambda_{g}}(1-x^{{3+\lambda_{g}}})+\frac{\lambda_{g}}{1+\lambda_{g}}(1-x^{{1+\lambda_{g}}})\nonumber\\
&&-\frac{2\lambda_{g}}{2+\lambda_{g}}(1-x^{{2+\lambda_{g}}}))+(2x^{{3+\lambda_{g}}}+x^{{1+\lambda_{g}}}-2x^{{2+\lambda_{g}}})]dt\nonumber\\
&&+\frac{1.852}{\pi^{2}}{\int_{t_{0}}^{t}}\alpha_{s}^{2}G(x,t)dt\hspace{4cm}
\end{eqnarray}
which defines the solution for $\lambda_{S}$. In this equation
$\lambda_{S_{0}}=\frac{{\partial}{\ln}F_{2}(x,t_{0})}{{\partial}{\ln}\frac{1}{x}}$
and  $F_{2}(x,t_{0})$ is input structure function at the starting
scale $t_{0}$. In calculations, we use the structure function
(Eq.24) and take the gluon distribution (Eq.16) for
singlet exponent.\\

\subsection{4. Discussion and Conclusions}

In this research we employed the Regge- like behavior of singlet
structure and gluon distribution function to calculate exponent
$\lambda_{S}$ singlet and $\lambda_{g}$ gluon based on NLO- DGLAP
evolution equations. Also, the singlet structure function and the
gluon distribution are calculated at the $x$ and $Q^{2}$ range of
HERA data. We compared our results with the experimental results
from $H1$ collaboration [3]. The results of calculations are shown
in Figs.1-8. For our calculation the
$\Lambda$ is equal to $292\hspace{0.1cm}$MeV corresponding to Ref.[3].\\

In Fig.1 we show $\lambda_{g}$ calculated from Eq.(18) as a
function of $x$ at four different fixed $Q^{2}$ values from 12 to
25 GeV$^{2}$. We observe that the derivative $\lambda_{g}$ is
almost independent of $x$ and consistent with the HERA data
 and compared with the MRST2001[17] fit. In Fig.2 we observe the
exponent rises almost linearly with $t(=\ln(Q^{2}/\Lambda^{2}))$.
This behavior is consistent with HERA data.\\

 In order to test the validity of our
obtained exponent gluon distribution, we calculate the gluon
distribution functions  using Eq.(17) and  compare them with the
theoretical predictions starting with the evolution at
$Q_{0}^{2}=5 \hspace{0.1cm}$GeV$^{2}$. As it can be seen in Fig.3,
  the values of $G(x,Q^{2})$ increase as
  $x$ decreases. Comparing these values with those of
  NLO-GRV  ${[15]}$, Donnachie $\&$ Landshoff (DL)[18] and MRST2001 [17] we can observe that
  these  gluon distribution function values are in agreement with the input parameterization at low
  $x$. In all cases the gluon distribution functions are increases
  as $Q^{2}$ is increased.\\

In Fig.4 we show $\lambda_{S}$ calculated from Eq.(26) as a
function of $x$ at four different fixed $Q^{2}$ values. we observe
at low $x$ the derivative $\lambda_{S}$ is almost independent of
$x$ consistent with the $H1$ [3] data explored in this range. This
implies that the $x$ dependence of $F_{2}$ at low $x$ is
consistent with a power law,
$F_{2}{\propto}\hspace{0.1cm}x^{-\lambda_{S}}$, for fixed $Q^{2}$.
Since quark contributions to the scaling violation of $F_{2}$ are
negligible , the low $x$ behavior is driven solely by the gluon
field. At larger $x$ the transition to the valence- quark region
causes a strong dependence of $\lambda_{S}$ on $x$ as indicated by
the NLO-QCD curves in Fig.1 of Ref.[3].\\

In Fig.5 we compare our predictions with $H1$ [3] data for
$\lambda_{S}$ as a function of $Q^{2}$ at six different fixed $x$
values. The derivative $\lambda_{S}$  of the proton singlet
structure function is observed to rise approximately
logarithmically with $Q^{2}$. It can be represented by a function
$\lambda_{S}(Q^{2})$ which is independent of $x$ within the
experimental accuracy. Of course, there is a mild $x$ dependency
within the experimental total errors. This is consistent with the
slopes extracted by Desgrolard \textit{et al} [19]. At this point
it is worth nothing that though the $x$- slope is often assumed to
be constant for small $x\hspace{0.1cm}{\leq}\hspace{0.1cm}0.01$,
there is no theoretical justification for a constant
$\lambda_{S}(Q^{2})$ at each fixed $Q^{2}$ . In fact there are
many models of structure functions that predict a varying
$\lambda_{S}(Q^{2})$ with $x$. For example, in the two Pomeron
model, $\lambda_{S}(Q^{2})$ has been shown to vary significantly
with $x$ [20]. Similarly in the generalized double asymptotic
scaling (DAS) [21], a tiny $x$ dependence is developed in the
effective slopes $\lambda_{S}(Q^{2})$ of the structure functions.
The Regge- based models [22] also predict a decrease of
$\lambda_{S}(Q^{2})$ for fixed $Q^{2}$ as
$x{\rightarrow}\hspace{0.1cm}0$.\\

The function $\lambda_{S}(Q^{2})$ is determined from fits of the
form $F_{2}(x,Q^{2})= A_{S}(Q^{2})x^{-\lambda_{S}}$ to  our
structure function data, restricted to the region
$x\hspace{0.1cm}{\leq}\hspace{0.1cm}0.01$. The results for
$A_{S}(Q^{2})$ and $\lambda_{S}(Q^{2})$ are presented in Figs.6
and 7. The coefficients $A_{S}(Q^{2})$ are approximately
independent of $Q^{2}$ with a mean value of $0.192$ (see Fig.7).
As can be seen in Fig.6, $\lambda_{S}(Q^{2})$ has an approximate
linear rise with ${\ln}Q^{2}$. It can be observed from these
figures that using Eq.(26) one can reach the good agreement with
the $H1$ Collaboration data [3] at low $x$. In Regge theory the
high energy behavior of hadron-hadron and photon-hadron total
cross section is determined by the pomeron intercept
$\alpha_{P}=1+\lambda$, and is given by
$\sigma_{\gamma(h)p}^{tot}(\nu){\sim}\nu^{\lambda}$. This behavior
is also valid for a virtual photon for $x<<1$, leading to the well
known behavior,$F_{2}{\sim}x^{-\lambda}$, of the structures at
fixed $Q^{2}$ and $x{\rightarrow}0$. The power $\lambda$ is found
to be either $\lambda{\sim}0.08$ or $\lambda{\sim}0.3$. The first
value corresponds to the soft Pomeron and the second value the
hard (Lipatov) Pomeron intercept[18].\\

However, due to the existence of absorptive corrections, this is
not the true Pomeron intercept (sometimes referred to as "bare"
Pomeron), but rather an effective one. The bare Pomeron intercept
is in fact substantially larger than $1+\lambda_{0}$. Indeed the
relative contribution of the most important absorptive corrections
decreases quite rapidly when $Q^{2}$ increases (like $Q^{-2}$ for
the eikonal ones ), so that as $Q^{2}{\rightarrow}\infty$ we see a
Pomeron intercept which is close to the bare Pomeron much larger
than the soft Pomeron. This consideration have prompted us to use
a low $x$ behavior of the structure function of the form:
\begin{equation}
F_{2}(x,Q^{2}){\sim}x^{-\lambda(Q^{2})} ;
\hspace{1cm}\lambda(Q^{2})=\lambda_{0}(1+\frac{2Q^{2}}{Q^{2}+d}),
\end{equation}
where $\lambda_{0}$ and $d$ are free parameters [23]. In Fig.6 we
observe this behavior.\\

The Form $x^{-\lambda_{g}}$ for the gluon parametrization at
 small $x$ is suggested by Regge behavior, but whereas the
 conventional Regge exchange is that of the soft Pomeron, with
$\lambda_{g}{\sim}0.0$, one may also allow for a hard Pomeron with
$\lambda_{g}{\sim}0.5$. The form $x^{-\lambda_{S}}$ in the sea
quark parametrization comes from similar considerations since, at
small $x$, the process $g{\rightarrow}q\overline{q}$ dominates the
evolution of the sea quarks. Hence the fits to early HERA data
have as a constraint $\lambda_{S}=\lambda_{g}$. However, one only
expects this once $Q^{2}$ is large enough for the effect of DGLAP
evolution to be seen, hence it may not be a reasonable constraint
at $Q^{2}=Q_{0}^{2}$. Furthermore, the exact solution of the DGLAP
equations predicts that $\lambda_{S}=\lambda_{g}-\epsilon$. The
data at low $x$ are now of sufficient precision that $\lambda_{S}$
and $\lambda_{g}$ to be separate free parameters, as in the MRST
fits. One notes that the gluon low $x$ slope has become
valence-like; however, this quickly changes as $Q^{2}$ increases,
such that $\lambda_{S}$ and $\lambda_{g}$ become equal at low
$Q^{2}$ and for larger $Q^{2}$, $\lambda_{g}{>}\lambda_{S}$, as
expected by PQCD [4,11].
 The evolution of the parameters $\lambda_{S}$ and $\lambda_{g}$ with $Q^{2}$  observe in Figs.2 and
 6. In these figures we see the differences between the
 values $\lambda_{S}{\sim}0.30$ and $\lambda_{g}{\sim}0.37$ are consistent to
 PQCD. This difference shows
that at low $x$ the gluon distribution is more singular than the
quark distribution. In other words the gluon is by far the most
dominant parton and $F_{2}$ is essentially given by the singlet
sea quark distribution which is driven by the gluon (through the
$g{\longrightarrow}q\overline{q}$
 splitting process) as $x{\rightarrow}0$. The small changes in the  quark
 distribution exponent
can be accompanied by large changes in the gluon distribution exponent.\\

In Fig.8 the calculation of the structure function
$F_{2}(x,Q^{2})$ at low $x$ is shown as a function of $x$. In this
figure we observe a continuous rise towards low $x$. From the
figures it can be seen that the results are well described  for
all $Q^{2}$- values by the NLO QCD fit, as is discussed in detail
in section 7.2 of Ref.[16]. The ${\ln}Q^{2}$ dependence of $F_{2}$
is observed to be non- linear. It can be well described by a
quadratic expression
\begin{equation}
F_{2}(x,Q^{2})=a(x)+b(x)lnQ^{2}+c(x)({\ln}Q^{2})^{2}.
\end{equation}
This equation is nearly coincides with the QCD fit in the
kinematics range of
these calculations.\\

The predictions for the small $x$ region measured at HERA shown in
Fig.8. It follows from this figure that the initial conditions,
 based on our knowledge of the Pomeron properties inferred
from soft and hard processes, together with conventional QCD
evolution, can explain the increase at very small $x$ observed at
HERA [3]. Our structure functions  compared with  the DL [10,20]
model, and Multipole Pomeron (MP) [24] with unit intercept. Here
Multipole Pomeron means that the Pomeron is a
multipole instead of just a simple pole.\\

The small- $x$ calculated data for the proton structure function
$F_{2}(x,Q^{2})$ show that a hard pomeron,  with intercept close
to 1.3, must be added to the familiar soft pomeron. So the
simplest fit to the small- $x$ data corresponds to
\begin{equation}
F_{2}(x,Q^{2})=\sum_{i=0,1}f_{i}(Q^{2})x^{-\epsilon_{i}},
\end{equation}
where the $i=0$ term is hard pomeron exchange and $i=1$ is soft
pomeron exchange. These parameters (i.e. $f_{i}(Q^{2})$ and
$\epsilon_{i}$) obtained from the best fit to all the small- $x$
data for $F_{2}(x,Q^{2})$ together with the data for
$\sigma^{{\gamma}p}$ [20]. Having concluded that the data for
$F_{2}$ require a hard pomeron component, it is necessary to test
this with our results. We compared our results with the two
pomeron fit as is seen in Fig.8. The agreement between our
calculated structure function, and its extraction of the hard
pomeron fit, is somewhat a striking success both of the
hard pomeron concept and of perturbative QCD.\\

To conclude, in this paper we have obtained a solution of the
DGLAP equation for the exponent $\lambda_{S}(x,Q^{2})$ and
$\lambda_{g}(x,Q^{2})$ in the next- to- leading order (NLO) at low
$x$. Our results show that of the derivatives
$\partial{{\ln}F_{2}(x,Q^{2})}/\partial{{\ln}1/x}{\equiv}\lambda_{S}(x,Q^{2})$
of the proton structure function $F_{2}(x,Q^{2})$ and
$\partial{{\ln}G(x,Q^{2})}/\partial{{\ln}1/x}{\equiv}\lambda_{S}(x,Q^{2})$
of the gluon distribution $G(x,Q^{2})$  are independent of $x$ for
$x{\leq}10^{-2}$ . We see that  $\lambda_{g}{>}\lambda_{S}$, as
expected by PQCD. We calculated  these quantities  as a function
of $Q^{2}$ at fixed $x$ and as a function of $x$ at fixed $Q^{2}$
consistent with the $H1$ data [3]. Thus the behavior of $F_{2}$
 at low $x$ is consistent with a dependence
 $F_{2}(x,Q^{2})=A_{S}x^{-\lambda_{S}}$  throughout that region.
 At low $x$, the exponent $\lambda_{S}$ is observed to rise
linearly
 with ${\ln}Q^{2}$ and the coefficient $A_{S}$ is independent of
 $Q^{2}$  within the experimental accuracy. This behavior of the structure function  $F_{2}(x,Q^{2})$ at low $x$ is consistent
 with a power- law behavior. Since at low $x$, $F_{2}(x,Q^{2})$
 is primarily driven by the gluon we also expect similar behavior for
 the gluon. The calculated slopes are consistent with
 experimental observations. But in the calculations we observed
  there is a mild $x$ dependency, consistent with the two
 pomeron model. Our results are compared with the DL and MP  modeles.\\
Also we calculated the structure function and the
 gluon distribution function at low $x$. We have compared our results with the QCD
parton distribution functions. Careful investigation of these
results show an agreement with the QCD gluon distributions. The
gluon distribution will increase as usual when $x$ decreases. Our
results suggest that evolutions both in $x$ and $Q^{2}$ have
solutions which have strong resemblances to BFKL- like behavior.\\
\newpage
\textbf{References}\\
1.Yu.L.Dokshitzer, Sov.Phys.JETP {\textbf{46}}, 641(1977);
G.Altarelli and G.Parisi, Nucl.Phys.B \textbf{126}, 298(1977);
V.N.Gribov and L.N.Lipatov,
Sov.J.Nucl.Phys. \textbf{15}, 438(1972).\\
 2. A.De Rujula \textit{et al}., phys.Rev.D \textbf{10}, 1649(1974);
 R.D.Ball and S.Forte, Phys.Lett.B \textbf{335}, 77(1994).\\
 3. $H1$ Collab., C.Adloff \textit{et al}., phys.Lett.B \textbf{520}, 183(2001).\\
 4. A.V.Kotikov and G.Parente, Phys.Lett.B \textbf{379}, 195(1996).\\
 5. J.K.Sarma and G.K.Medhi, Eur.Phys.J.C \textbf{16}, 481(2000).\\
 6. P.D.Collins, \textit{An introduction to Regge theory an
high-energy physics}(Cambridge University Press, Cambridge 1997)Cambridge.\\
7. M.Bertini \textit{et al}., Rivista del Nuovo Cimento \textbf{19}, 1(1996).\\
8. A.V.Kotikov and G.Parente, Phys.Lett.B \textbf{379}, 195(1996).\\
9. E.A.Kuraev, L.N.Lipatov and V.S.Fadin, Phys.Lett.B \textbf{60},
50(1975); Sov.Phys.JETP \textbf{44}, 433(1976); ibid. \textbf{45},
199(1977);
Ya.Ya.Balitsky and L.N.Lipatov, Sov.J.Nucl.Phys. \textbf{28}, 822(1978).\\
10. A.Donnachie and P.V.Landshoff, Z.Phys.C \textbf{61},
139(1994);
Phys.Lett.B \textbf{518}, 63(2001).\\
11.A.M.Cooper- Sarkar and R.C.E.D Evenish,
Acta.Phys.Polon.B \textbf{34}, 2911(2003).\\
12.W.Furmanski and R.Petronzio, Phys.Lett.B \textbf{97},
437(1980);
Z.Phys.C \textbf{11}, 293 (1982).\\
13. R.K.Ellis, Z.Kunszt and E.M.Levin, Nucl.Phys.B \textbf{420},
517(1994).\\
14. R.K.Ellis , W.J.Stirling and B.R.Webber, \textit{QCD and
Collider Physics}(Cambridge University
Press)1996.\\
15.M.Gluk, E.Reya and A.Vogt,Z.Phys.C \textbf{67}, 433(1995);
 Eur.Phys.J.C \textbf{5}, 461(1998).\\
16. $H{1}$ Collab., C.Adloff \textit{et al}., Eur.Phys.J.C \textbf{21}, 33(2001).\\
17.A.D.Martin, R.G.Roberts, W.J.Stirling and R.S.Thorne, Phys.Lett.B \textbf{531}, 216(2002).\\
18.A.Donnachie and P.V.Landshoff, Phys.Lett.B \textbf{550},
160(2002); R.D.Ball and P.V.Landshoff, arXiv:hep-ph/9912445.\\
19.P.Desgrolard, A.Lengyel and E.Martynov, J.High Energy
Phys. \textbf{0202}, 029(2002).\\
20. P.V.Landshoff, hep-ph/0203084.\\
21. A.V.Kotikov and G.Parente, Nucl.Phys.B \textbf{549},
242(1999);
hep-ph/0207276.\\
22. P.Desgrolard, A.Lengyel and E.Martynov, Eur.Phys.J.C
\textbf{7}, 655(1999); P.Desgrolard and E.Martynov, Eur.Phys.J.C
\textbf{22},
479(2001); J.R.Cudell and G.Soyez, Phys.Lett.B \textbf{516}, 77(2001).\\
23.A.Capella \textit{et al}., Phys.Lett.B \textbf{337},358(1994);
Phys.Rev.D \textbf{63},
054010(2001).\\
24.L.Csernai \textit{et al}., hep-ph/0112265.\\
\newpage
\textbf{Figure Captions}\\
Fig.1. Exponents $\lambda_{g}$ plotted against $x$ at four fixed
$Q^{2}$ values and compared with
$\lambda_{g}={\partial}{\ln}G(x,Q^{2})/{\partial}{\ln}(1/x)$ of
MRST2001[17] fit.\\

Fig.2. Calculation of the exponent $\lambda_{g}$ from fits of the
form $G(x,Q^{2})=\hspace{0.1cm} A_{g}x^{-\lambda_{g}}$ to the
our gluon distribution data for $x{\leq}\hspace{0.1cm}0.01$.\\

Fig.3. The gluon distribution given by Eq.(17) against $x$ at four
fixed $Q^{2}$ values and compared with NLO-GRV[15](Solid
line) , DL fit[18](Dash line) and MRST fit[17](Dot line).\\

Fig.4. Exponents $\lambda_{S}$ plotted against $x$ at four fixed
$Q^{2}$ values and compared with data from $H1$ [3]. The error
bars represent the statistical and systematic uncertainties added
in quadrature.\\

Fig.5. Exponents $\lambda_{S}$ plotted against $Q^{2}$ at
different fixed $x$ values and compared with data from $H1$ [3].
The error bars represent the statistical and systematic
uncertainties added in quadrature.\\

Fig.6. Calculation of the exponent $\lambda_{S}$ from fits of the
form $F_{2}(x,Q^{2})=\hspace{0.1cm} A_{S}x^{-\lambda_{S}}$ to the
our structure function  data for $x{\leq}\hspace{0.1cm}0.01$ and
compared with data from $H1$ [3]. The error bars represent the
statistical and systematic uncertainties added in quadrature .
Solid line is the Capella [23] fit with the $^{,,}$bare$^{,,}$
Pomeron intercept.\\

 Fig.7. Calculation of the coefficient
$A_{S}(Q^{2})$ from fits of the form
$F_{2}(x,Q^{2})=\hspace{0.1cm} A_{S}x^{-\lambda_{S}}$ to the our
structure function  data for $x{\leq}\hspace{0.1cm}0.01$ and
compared with data from $H1$[3]. The error bars represent the
statistical and systematic uncertainties added in quadrature.\\

Fig.8. The calculated values of the structure function
$F_{2}(x,Q^{2})$ plotted as functions of $x$ in our method,
compared with  NLO QCD fit to the $H1$ data [16](solid line), two-
pomeron fit [20](dash line), and  Multipole Pomeron exchange fit
(MP model)[24].\\
\newpage
\begin{figure}
\centering
\includegraphics[width=1\textwidth]{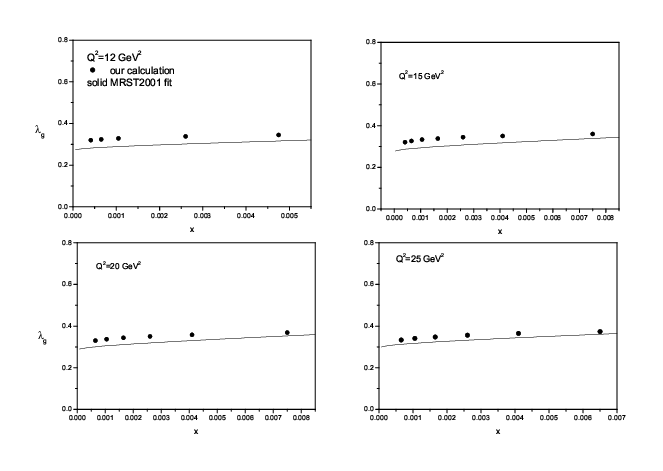}
Fig.1
\end{figure}
\begin{figure}
\centering
\includegraphics[width=1\textwidth]{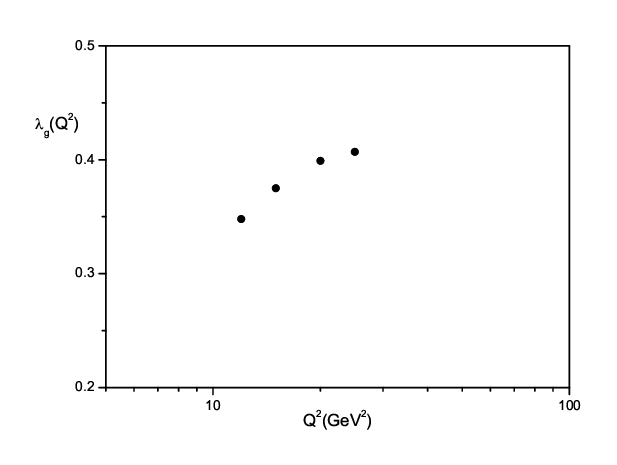}
Fig.2
\end{figure}
\begin{figure}
\centering
\includegraphics[width=1\textwidth]{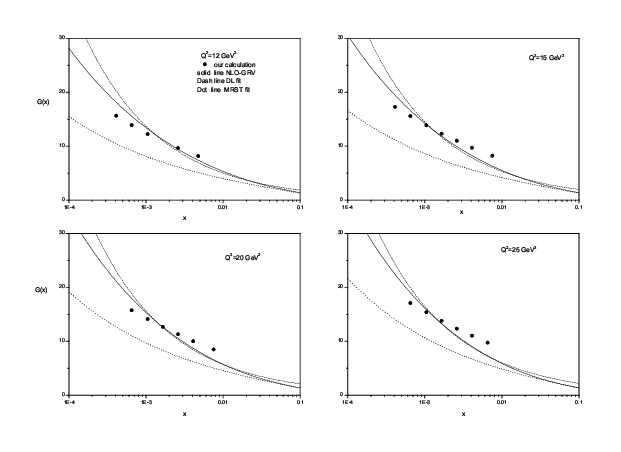}
Fig.3
\end{figure}
\begin{figure}
\centering
\includegraphics[width=1\textwidth]{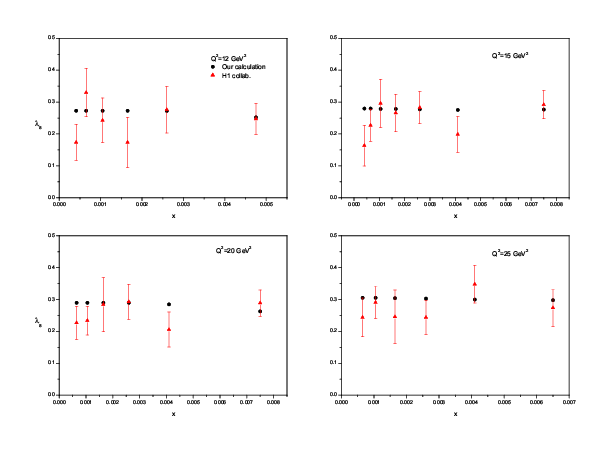}
Fig.4
\end{figure}
\begin{figure}
\centering
\includegraphics[width=1\textwidth]{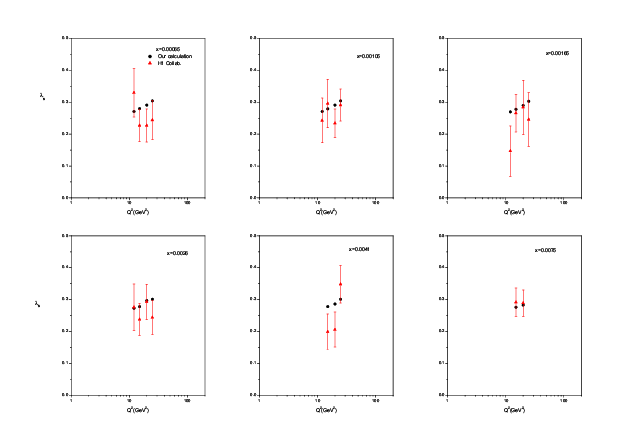}
Fig.5
\end{figure}
\begin{figure}
\centering
\includegraphics[width=1\textwidth]{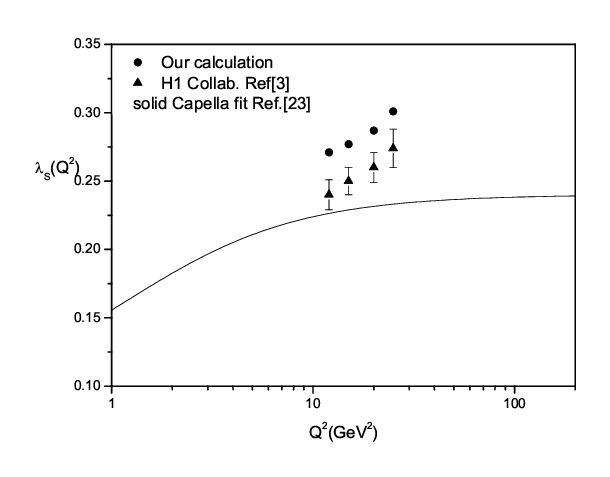}
Fig.6
\end{figure}
\begin{figure}
\includegraphics[width=1\textwidth]{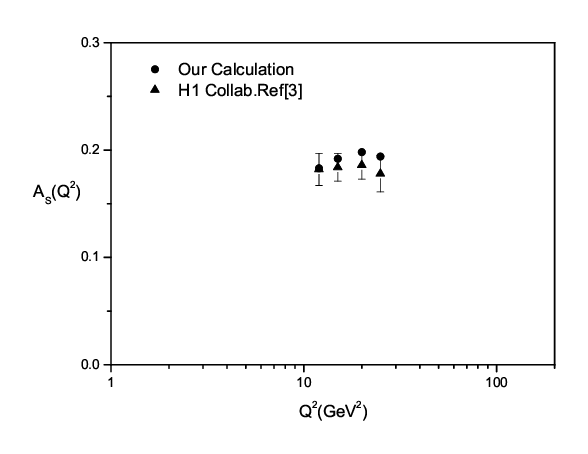}
Fig.7
\end{figure}
\begin{figure}
\centering
\includegraphics[width=1\textwidth]{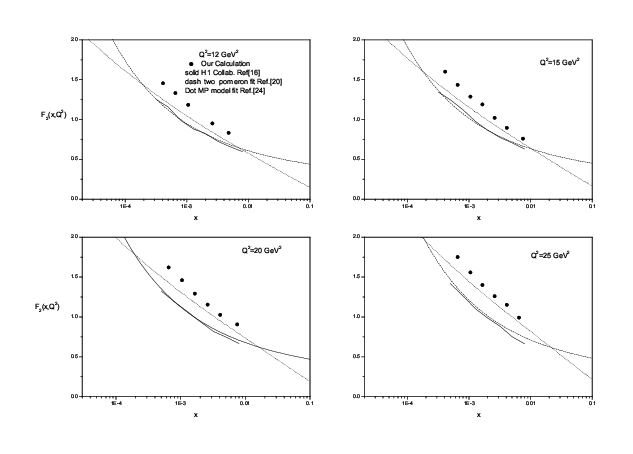}
Fig.8
\end{figure}
\end{document}